# Higher-magnesium-doping effects on the singlet ground state of the Shastry-Sutherland SrCu$_2$(BO$_3$)$_2$

Lia Šibav[1,2], Žiga Gosar[1,3], Tilen Knaflič[1,4], Zvonko Jagličić[5,6], Graham King[7], Hiroyuki Nojiri[8], Denis Arčon[1,3] and Mirela Dragomir[1,2*]

[1]Jožef Stefan Institute, Jamova cesta 39, 1000 Ljubljana, Slovenia
[2]Jožef Stefan International Postgraduate School, Jamova cesta 39, 1000 Ljubljana, Slovenia
[3]Faculty of Mathematics and Physics, University of Ljubljana, Jadranska ulica 19, 1000 Ljubljana, Slovenia
[4]Institute for the Protection of Cultural Heritage of Slovenia, Research Institute, Poljanska cesta 40, 1000 Ljubljana, Slovenia
[5]Institute of Mathematics, Physics and Mechanics, Jadranska ulica 19, 1000 Ljubljana, Slovenia
[6]Faculty of Civil and Geodetic Engineering, University of Ljubljana, Jamova cesta 2, 1000 Ljubljana, Slovenia
[7]Canadian Light Source, 44 Innovation Blvd, Saskatoon, SK S7N 2V3, Canada
[8]Institute for Materials Research, Tohoku University, Katahira 2-1-1, Sendai, 980-8577 Japan

*Corresponding author: mirela.dragomir@ijs.si

**Abstract**

Doping of quantum antiferromagnets is an established approach to investigate the robustness of their ground state against the competing phases. Predictions of doping effects on the ground state of the Shastry-Sutherland dimer model are here verified experimentally on Mg-doped SrCu$_2$(BO$_3$)$_2$. A partial incorporation of Mg$^{2+}$ on the Cu$^{2+}$-site in the SrCu$_2$(BO$_3$)$_2$ structure leads to a subtle but systematic lattice expansion with the increasing Mg-doping concentration, which is accompanied by a concomitant decrease in the spin gap, the Curie-Weiss temperature and the peak temperature of the susceptibility. These findings indicate a doping-induced breaking of Cu$^{2+}$ spin-1/2 dimers which is also corroborated by X-band EPR spectroscopy that points to a systematic increase in intensity of free Cu$^{2+}$ sites with increasing Mg-doping concentration. Extending the Mg-doping up to nominal $x$ = 0.10 or SrCu$_{1.9}$Mg$_{0.1}$(BO$_3$)$_2$, in the magnetisation measurements taken up to 35 T, a suppression of the pseudo-1/8 plateau is found along with a clear presence of an anomaly at an onset critical field $H'_{C0} \approx 9$ T. The latter, absent in pure SrCu$_2$(BO$_3$)$_2$, emerges due to the coupling of liberated Cu$^{2+}$ spin-1/2 entities in the vicinity of Mg-doping induced impurities.

## 1. INTRODUCTION

Quantum fluctuations, low dimensionality and geometric frustration are considered important ingredients for exotic states of matter such as quantum spin liquids [1,2] or high-$T_C$ superconductivity [3]. In particular, low-dimensional materials with antiferromagnetic (AFM) interactions and coupled $S$ = ½ moments are the most promising to experimentally explore the realisation of such exotic states [4,5,6,7]. In the last decades, transition metal oxides exhibiting such physics have been actively studied. A particular interest is devoted to the family of high-$T_C$ cuprate superconductors where the observation of two-dimensional (2D) antiferromagnetism in Mott-insulating parent compounds led to the suggestion of an intimate correlation of the gapped spin excitations with unconventional superconductivity in charge-doped compounds [8]. Following this discovery, numerous studies focused on other 2D spin systems with spin-gapped ground states [9,10]. While the spin-singlet ground state is commonly found in one-dimensional (1D) or quasi-1D materials [11,12,13,14,15], there are only few examples of 2D spin-gapped systems [16,17,18] due to their tendency of long-range ordering guided by the interlayer exchange interactions.

One rare example of a spin-gapped 2D material is strontium copper orthoborate, $SrCu_2(BO_3)_2$ [18]. As the first experimental realisation of the theoretical Shastry-Sutherland model [19], this frustrated 2D spin system contains spin dimers which are orthogonally coupled on a square lattice. In this model, the in-plane coupling of $Cu^{2+}$ ($S = 1/2$) spins is described by the antiferromagnetic nearest neighbour (*NN*) coupling $J$ and the next-nearest neighbour (*NNN*) coupling $J'$:

$$\hat{H} = J \sum_{NN} S_i S_j + J' \sum_{NNN} S_i S_j + g\mu_B B \cdot \sum_i S_i . \qquad (1)$$

The last term in **Equation 1** is the Zeeman interaction with the external magnetic field, $B$. The model Hamiltonian has been successfully applied to $SrCu_2(BO_3)_2$ to interpret, e.g., the results of electron paramagnetic resonance (EPR) [20,21,22,23,24,25] and nuclear magnetic resonance (NMR) studies [26,27,28,29]. Importantly, the $J'/J$ ratio in **Equation 1** dictates the magnetic ground state [19,30]. For $J'/J < 0.69$, the ground state is the exact dimer phase that is a product of singlets on Cu-dimers [31], while a gapped plaquette singlet state or a plaquette valence bond solid (PVBS) phase is found for $0.69 < J'/J < 0.86$ [30,31]. For $J'/J > 0.86$, the Néel phase or the dimer valence bond solid (DVBS) phase is stabilised. Recently, a novel quantum spin liquid phase has been predicted to exist between the gapped plaquette-singlet phase and the antiferromagnetic phase [32]. Experimentally, the $J'/J$ ratio in $SrCu_2(BO_3)_2$ was determined to be 0.68 ($J$ = 100 K, $J'$ = 68 K) [33], with more recent reports claiming a slightly smaller value of 0.63(1) [34,35]. The value of $J'/J$ positions $SrCu_2(BO_3)_2$ to the exact singlet-dimer state of the Shastry-Sutherland phase diagram. However, it is also very close to the quantum critical point that separates the singlet-dimer state to other competing phases [28]. This specific property of $SrCu_2(BO_3)_2$ opens up an intriguing possibility to investigate the stability and behaviour of a frustrated quantum antiferromagnet close to a quantum critical point.

The exact dimer phase of $SrCu_2(BO_3)_2$ was first challenged in high-magnetic field experiments aiming to close the spin gap, $\Delta$. The first estimate of a spin gap $\Delta$ of 19(1) K was from low-temperature magnetic susceptibility [36], but more precise values of 34(1) K were obtained from other experiments such as high-field EPR [37]. Upon closing $\Delta$ with magnetic field, the presence of strong frustration is responsible for the Wigner-like crystallisation of triplets at critical fields, giving rise to characteristic magnetisation plateaus [38]. Namely, the magnetisation plateaus at fractional values of the saturated magnetisation of 1/8, and 1/4 were initially observed [18,36]. Later, other intermediate fractional plateaus were also reported such as 1/3 [39] and 1/9, 1/7, 1/6, 1/5, 2/9 [40], adding to the extreme richness of the $SrCu_2(BO_3)_2$ phase diagram. A very recent study showed evidence of a spin nematic phase or bound-state condensate in $SrCu_2(BO_3)_2$ before the 1/8 plateau and beyond 21 T, which is understood as a condensate of bosonic Cooper pairs [41].

Alternatively, the ground state of $SrCu_2(BO_3)_2$ may be altered by means of physical pressure or chemical pressure through substitutions. Applying hydrostatic pressure on $SrCu_2(BO_3)_2$ [27,42,43] was shown to induce two quantum phase transitions: i) around 1.7 GPa, where the dimer singlet phase transitions to a plaquette singlet phase below 2 K, ii) at 3–4 GPa, where another antiferromagnetic phase is realised below 4 K [44].

On the other hand, the effects of chemical pressure or chemical-doping-induced impurities on the ground state of $SrCu_2(BO_3)_2$ are less understood due to difficulties in chemical modifications of this compound. However, theoretical studies have predicted that altering the ground state of $SrCu_2(BO_3)_2$ by chemical doping could result in exotic magnetic states such as quantum spin liquids or superconductors [45,46]. In spite of these promising theoretical predictions, the literature on the experimental doping of $SrCu_2(BO_3)_2$ is scarce [47,48,49,50,51,52] as introducing dopants in the crystal structure of $SrCu_2(BO_3)_2$ has been proven to be a very challenging process even for low doping concentrations.

The chemical substitutions on the Sr site, located between the Cu–O–B layers in the $SrCu_2(BO_3)_2$ structure, could exert both chemical pressure on the Cu–Cu exchange pathways and charge imbalance through a different valence state of the dopant, affecting the Cu magnetic moment. Doping on the Sr site with Y, Na, La, Zn or Al led to a suppression of the spin gap in polycrystalline samples [49], but the gap was not fully closed and no superconductivity was observed. Single crystals of Na- and La-doped

SrCu$_2$(BO$_3$)$_2$ have also been grown by the optical floating zone method [50]. Although the XRD measurements showed that these compounds are single-phase and their in-plane lattice parameter depends systematically on the dopant content $x$, a detailed structural and local-probe investigation such as EPR or NMR of these dopants is still missing.

Moreover, a quantum spin liquid is anticipated to be realised by magnetic dilution – partially substituting the magnetic Cu$^{2+}$ ions with non-magnetic isovalent cations, which could break the spin dimers and establish a long-range quantum entanglement at the lowest temperatures [47].

Magnetic dilution, i.e., Mg$^{2+}$ substitutions on the Cu$^{2+}$ site, has been proven to be chemically even more challenging and only a few reports exist on both polycrystalline and single crystals of Mg-doped SrCu$_2$(BO$_3$)$_2$. In the first report, nominal 2.5% and 5% Mg-doped SrCu$_2$(BO$_3$)$_2$ single crystals were grown by optical floating zone method [50]. These samples were further used for a series of inelastic neutron scattering experiments where they observed both triplet excitations and in-gap excitations [51]. As the ionic radii of Cu$^{2+}$ and Mg$^{2+}$ are almost identical (~0.57 Å) for a square planar coordination [53], the observed subtle and non-systematic structural changes are not surprising. This renders also the determination of the true range of doping difficult. However, while Cu$^{2+}$ can adopt both an octahedral and a square planar coordination, as in SrCu$_2$(BO$_3$)$_2$, Mg$^{2+}$ chemically prefers an octahedral coordination. Therefore, achieving a substantial concentration of Mg incorporated into the SrCu$_2$(BO$_3$)$_2$ structure still remains a challenge.

With the aim of pushing the doping level to higher values and addressing the robustness of the exact dimer phase, this paper reports a comprehensive study of Mg-doping of polycrystalline SrCu$_2$(BO$_3$)$_2$, targeting the 10 mol% nominal doping concentration or $x$ = 0.10 in SrCu$_{2-x}$Mg$_x$(BO$_3$)$_2$. To achieve such high doping levels, a different source of magnesium was used, namely Mg$_2$CO$_3$(OH)$_2 \cdot$ 3.75 H$_2$O, which is expected to increase the reactivity and help with Mg-incorporation. A successful substitution of Cu with Mg in the parent SrCu$_2$(BO$_3$)$_2$ structure is proven by powder X-ray diffraction and energy-dispersive X-ray spectroscopy. It is then shown that Mg-doping leads to a softening of the magnetism in SrCu$_2$(BO$_3$)$_2$ as suggested by a systematic decrease in the spin gap, $\Delta$, the Curie-Weiss temperature, $\theta$, and the maximum of the magnetic susceptibility, $T_{max}$. The effects of Mg-doping on the Cu$^{2+}$ dimers are further evaluated with EPR spectroscopy. These measurements show that the EPR spectra broaden with Mg-doping due to the increase in the concentration of intrinsic paramagnetic dimer-free Cu$^{2+}$ sites that are coupled to the spin dimers. Finally, the weak anomaly at an onset critical field of $H'_{c0} \approx 9$ T, previously first observed by Shi *et al.* [48] for their $x$ = 0.05 Mg-doped SrCu$_{2-x}$Mg$_x$(BO$_3$)$_2$, is here investigated using high-field magnetisation measurements up to 35 T. We find that this anomaly is more pronounced for $x$ = 0.10 and its intensity systematically increases with Mg concentration. This work thus constitutes a systematic study of the effect of magnetic dilution on the magnetic properties of SrCu$_2$(BO$_3$)$_2$ and opens alternative pathways towards higher substitutional doping of SrCu$_2$(BO$_3$)$_2$.

## 2. EXPERIMENTAL

### 2.1. Materials and Methods

Pristine and Mg-doped SrCu$_2$(BO$_3$)$_2$ polycrystalline samples SrCu$_{2-x}$Mg$_x$(BO$_3$)$_2$ with $x$ being 0.03, 0.05, and 0.1, respectively, were synthesised by the traditional solid-state method and using boric acid as the boron source. The overall reactions are summarised in **Equations 2** and **3**:

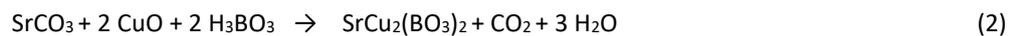

$$\text{SrCO}_3 + 2\,\text{CuO} + 2\,\text{H}_3\text{BO}_3 \rightarrow \text{SrCu}_2(\text{BO}_3)_2 + \text{CO}_2 + 3\,\text{H}_2\text{O} \qquad (2)$$

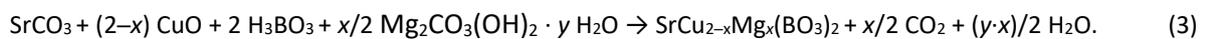

$$\text{SrCO}_3 + (2-x)\,\text{CuO} + 2\,\text{H}_3\text{BO}_3 + x/2\,\text{Mg}_2\text{CO}_3(\text{OH})_2 \cdot y\,\text{H}_2\text{O} \rightarrow \text{SrCu}_{2-x}\text{Mg}_x(\text{BO}_3)_2 + x/2\,\text{CO}_2 + (y \cdot x)/2\,\text{H}_2\text{O}. \qquad (3)$$

The starting materials were high-purity SrCO$_3$ (Alfa Aesar, 99.994%), CuO (Aldrich, 99.99%) and H$_3$BO$_3$ (Alfa Aesar, 99.9995%). The starting material used as Mg-source was Mg$_2$CO$_3$(OH)$_2 \cdot y$H$_2$O ($y \approx 3$) (Alfa Aesar, 99.996%). The coefficient $y$ in Mg$_2$CO$_3$(OH)$_2 \cdot y$H$_2$O ($y \approx 3$) was determined to be 3.75 by thermal gravimetric analysis. This precursor was stored and weighed in a MBraun glove box under an argon atmosphere of < 0.1 ppm $O_2$ and < 0.1 ppm $H_2O$.

The starting materials were weighed in stoichiometric ratios and hand-homogenised in an agate mortar for an hour. The homogenised grey powder was then pressed into pellets with an 8 or 10 mm diameter with a force of 5 tonnes. The pellets were fired for multiple cycles with several intermediate grindings, following the thermal profile: 780 °C for 24 h and 810 °C for 24 h in air plus multiple 3- or 6-day cycles at 900 °C in $O_2$ atmosphere, until no changes were noticed in the PXRD pattern. More details about this synthesis procedure can be found in [54].

### 2.2. Characterisation

**X-ray powder diffraction (PXRD)**

The phase composition of the obtained samples was investigated by powder X-ray diffraction (PXRD) using a PANalytical X'Pert Pro powder diffractometer with Cu-K$\alpha_1$ radiation in 2θ range 10–120° with a step size of 0.016° and a counting time of 300 s per step.

Synchrotron powder X-ray diffraction data were also collected at the Brockhouse high energy wiggler beamline at the Canadian Light Source (CLS) using an area detector of a 200 x 200 µm pixel size, and $\lambda$ = 0.3502 Å radiation with Ni as calibrant. The powders were packed into kapton capillaries and measured in 2θ range 1–26°. The structural refinements were performed with the Rietveld method using the program GSAS-II [55].

**Scanning Electron Microscopy (SEM) and Energy-dispersive X-ray spectroscopy (EDS)**

For the microstructural investigations, Mg-doped $SrCu_2(BO_3)_2$ powders were embedded in an epoxy resin matrix polished and carbon coated using a sputter coater model Balzers SCD 050. The imaging and compositional analyses were performed using Thermo Fisher Quanta 650 ESEM equipped with an energy-dispersive X-ray spectrometer (Oxford Instruments, AZtec Live, Ultim Max SDD 65 mm$^2$) and a field-emission-gun scanning electron microscope (FE-SEM; JEOL JSM-7600) further equipped with an energy-dispersive X-ray spectrometer (EDS; INCA Oxford 350 EDS SDD), electron backscatter diffraction (EBSD) and a wavelength-dispersive X-ray spectrometer (WDXS; INCAWave 500). The accelerating voltage was 20 kV.

**Magnetic susceptibility**

The magnetic susceptibility measurements were performed on an MPMS 3XL-5 SQUID magnetometer from Quantum Design within the 2–300 K temperature range and an applied magnetic field of 1 kG. Powder samples were placed inside a plastic capsule, which was inserted into a standard straw sample-holder and closed with a plastic stopper at one end.

**Electron paramagnetic resonance spectroscopy (EPR)**

The continuous wave (CW) X-band electron paramagnetic resonance spectroscopy experiments were performed on a Bruker E500 spectrometer operating at 9.37 GHz, equipped with a Varian TEM104 dual cavity resonator, an Oxford Instruments ESR900 cryostat and an Oxford Instruments ITC503 temperature controller. Samples were measured at temperatures ranging from room temperature down to 4 K, typically applying 2 mW of microwave power and a 5 G modulation amplitude with the modulation frequency of 50 kHz. In a typical experiment, approximately 30 mg of powder sample, sealed in a standard 4 mm EPR quartz tubes under dynamic vacuum, was used for each measurement.

**High-field magnetic susceptibility**

About 50 mg of polycrystalline Mg-doped $SrCu_2(BO_3)_2$ was inserted into kapton tubes, which were closed with stoppers from both sides. High-field magnetisation measurements were performed at 0.4 K in pulsed magnetic fields up to 35 T at Institute for Materials Research, Tohoku University.

## 3. RESULTS AND DISCUSSION

### 3.1. Powder X-ray diffraction (PXRD)

At room temperature, $SrCu_2(BO_3)_2$ crystallises in the $I\bar{4}2m$ space group – first reported in 1991 by Smith and Keszler [56], where distorted planar $CuO_4$ units and rigid $BO_3$ triangular groups form two-dimensional puckered layers of $[Cu(BO_3)_2]^{2-}$ units at $z \approx ¼$ stacked along the [001] direction. These layers are separated by non-magnetic $Sr^{2+}$ ions located at $z = 0, ½$ – **Figure 1a**. The distortion from planar $CuO_4$ rectangles is a consequence of sharing an O−O edge with a B atom while the other O–O edge is shared with a Cu atom. As a result of this distortion, the $[Cu(BO_3)_2]^{2-}$ layers are corrugated. The Cu–O distances measured along the $c$-axis are larger than 3.25 Å. All the in-plane $Cu^{2+}$ sites ($S = ½$) are crystallographically equivalent and a two-dimensional network of copper dimers is formed with the neighbouring pairs oriented orthogonal to each other – **Figure 1b**. The first nearest neighbour (NN) $Cu^{2+}$ pairs are distanced by 2.903 Å while the next-nearest neighbour (NNN) distance is at 5.133 Å.

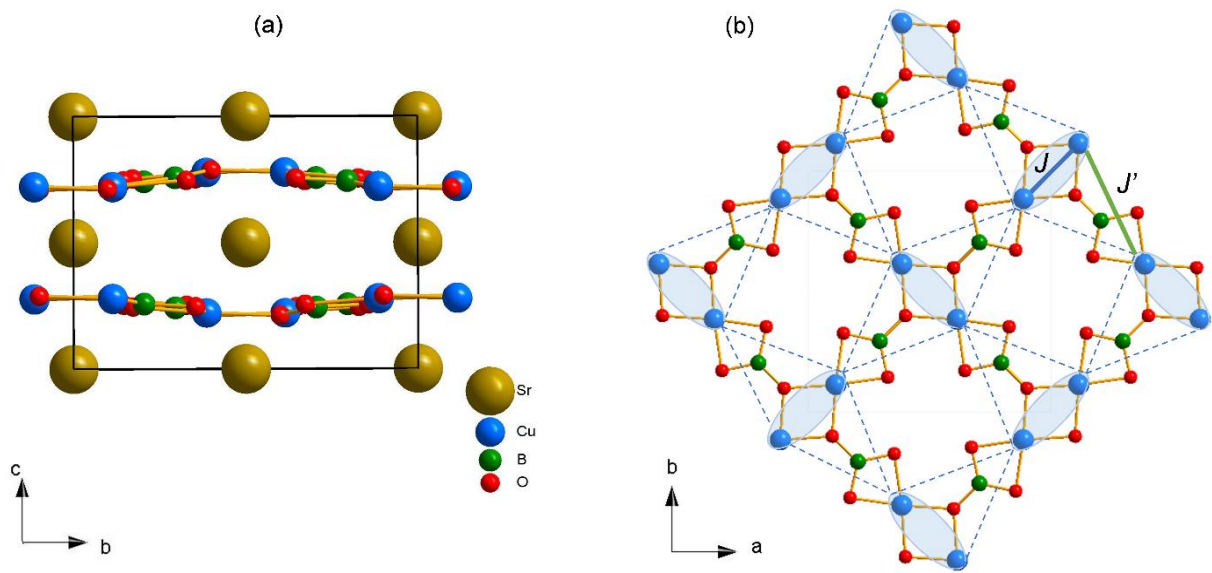

**Figure 1 (a)** The $SrCu_2(BO_3)_2$ structure viewed along the $c$-axis. The Cu–B−O layers can be seen separated by $Sr^{2+}$ ions. **(b)** View perpendicular to the Cu–B–O layer which shows the orthogonal $Cu^{2+}$ spins coupled across the dimer bonds with the NN coupling $J$, represented by the blue solid line, and across the next-neighbour bonds with the NNN coupling $J'$, represented by the green solid line.

High-resolution X-ray powder diffractogram of the pristine $SrCu_2(BO_3)_2$ shows the presence of the main $SrCu_2(BO_3)_2$ phase, with traces of CuO – **Figure 2a**. We note that the minute presence of the persistent CuO impurity could not be avoided in powder samples, which is in agreement with previous reports [50, 56]. The Rietveld refinement analysis of the powder diffraction data yields $a$ = 8.9900(2) Å and $c$ = 6.6483(2) Å. These parameters are in very good agreement with the values reported in the literature, e.g., $a$ = 8.998(1) Å and $c$ = 6.654(1) Å from [50] versus $a$ = 8.991(1) Å and $c$ = 6.660(3) Å in [56] or $a$ = 8.995(1) Å and $c$ = 6.649(1) Å reported in [49,52].

Upon Mg-doping, the same two phases are consistently observed: the majority of the $SrCu_2(BO_3)_2$ phase accompanied by the minor CuO impurity. However, the presence of CuO impurity grows with the nominal Mg concentration. Magnesium-based secondary phases are also present in the samples, as detected later with EDS analysis, with amounts increasing with the Mg concentration – **Figure 2a**.

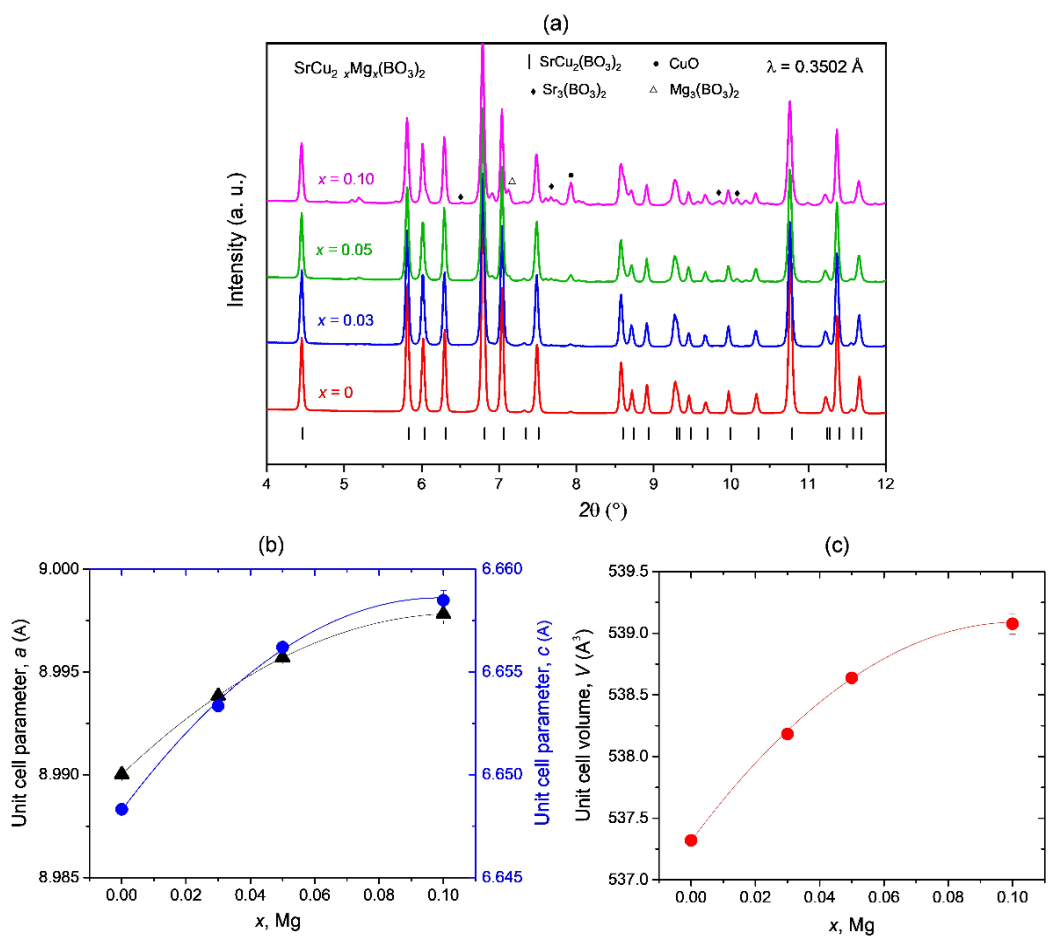

**Figure 2 (a)** A selected region of the room-temperature synchrotron PXRD patterns of $SrCu_{2-x}Mg_x(BO_3)_2$ with $x$ = 0.03, 0.05 and 0.10, respectively. The formation of the dominant tetragonal $SrCu_2(BO_3)_2$ phase can be seen in all patterns. Some CuO impurity (marked with solid circles), Mg-based secondary phases (open triangles) and $Sr_3(BO_3)_2$ (solid diamonds) were also detected in the doped samples. A systematic increase in the unit cell parameters **(b)** and the unit cell volume **(c)** with Mg-doping is observed.

The presence of these impurities underlines difficulties to incorporate Mg into the square planar coordination of Cu in the parent $SrCu_2(BO_3)_2$ structure and suggests that the actual Mg-doping values are lower than the nominal values. Nevertheless, the current synthesis approach leads to Mg incorporation judging from a monotonic shift of the diffraction peaks towards lower 2θ values and a systematic increase of the unit cell parameters and volume with the increasing Mg-doping concentration – **Figure 2b,c**. The unit cell parameters and unit cell volumes for $SrCu_{2-x}Mg_x(BO_3)_2$ with different $x$ as obtained from Rietveld refinement analysis are directly compared in **Supplementary File, Table S1**. Up to nominal $x$ = 0.05, both $a$ and $c$ lattice parameters show a steady increase with Mg concentration, but then tend to saturate towards nominal $x$ = 0.10. Specifically, the refined room temperature unit cell parameters increase to $a$ = 8.9938(2) Å and $c$ = 6.6534(2) Å for $x$ = 0.03, to $a$ = 8.9957(2) Å and $c$ = 6.6562(2) for $x$ = 0.05 and up to $a$ = 8.9978(5) Å and $c$ = 6.6584(5) Å for $x$ = 0.10, respectively – **Figure 2b**. Comparing these unit cell parameters with the available literature data for $x$

= 0.05, showing $a$ = 8.994(1), and $c$ = 6.648(1) [50], this study shows a slightly larger Mg-doping induced lattice expansion for this doping concentration. However, the difference is still within the 3% tolerance of the unit cell volume regularly reported for undoped $SrCu_2(BO_3)_2$ [49, 52, 56]. Compared to the parent pristine sample, the largest unit cell volume expansion of 0.33% is for the $x$ = 0.10. This increase is small and reflects the similarity in the ionic radius of $Cu^{2+}$ and $Mg^{2+}$ ions as well as the limited amount of the incorporated $Mg^{2+}$. When comparing $x$ = 0.05 and $x$ = 0.10 samples, the relative change in both lattice parameters is extremely small and it amounts only about 0.002 Å. This suggests that the synthesis approach used in this study may have an upper limit for the effective Mg-doping into the $SrCu_2(BO_3)_2$ structure up to these values.

A closer look at the structural details further reveals a systematic decrease in the Cu−Cu distance between $Cu^{2+}$ pairs from 2.9191(11) Å for pristine $SrCu_2(BO_3)$ ($x$ = 0) to 2.9091(15) Å for $SrCu_{1.95}Mg_{0.05}(BO_3)_2$ ($x$ = 0.05) and 2.8982(26) Å for $SrCu_{1.9}Mg_{0.1}(BO_3)_2$ ($x$ = 0.10). A shorter Cu−Cu bond length (blue line labelled *J* in **Figure 1b**) should also reduce the Cu−O−Cu bond angle. The Cu−O−Cu bond angle indeed slightly decreases with Mg doping, from 99.403(436)° for the undoped sample ($x$ = 0) to 99.049(571)° for $x$ = 0.05. As the Cu–O–Cu bond angle plays a substantial role in defining the effective superexchange interaction in strongly correlated transition metal oxides [57], this decrease in the bond angle tuned by the chemical pressure in the Mg-doped samples may lead to a small decrease in the strength of the intradimer antiferromagnetic exchange *J*. Meanwhile, a systematic increase in the interdimer distance Cu−Cu between the orthogonal dimers (green line labelled *J'* in **Figure 1b**) from 5.1224(14) Å for $SrCu_2(BO_3)_2$ ($x$ = 0) to 5.1297(12) Å for $SrCu_{1.95}Mg_{0.05}(BO_3)_2$ ($x$ = 0.05) and 5.1361(20) Å for $SrCu_{1.9}Mg_{0.1}(BO_3)_2$ ($x$ = 0.10) is observed. This is consistent with the overall lattice expansion – **Figure 2c**.

### 3.2. Scanning Electron Microscopy (SEM) and Energy-Dispersive X-ray spectroscopy (EDS)

The PXRD data clearly shows that the substitution of $Cu^{2+}$ with $Mg^{2+}$ in $SrCu_2(BO_3)_2$ is possible, although not fully up to the nominal Mg concentrations. However, a detailed chemical analysis is mandatory to unambiguously confirm the incorporation of Mg and to obtain more quantitative information on the dopant concentration and distribution within the matrix as well as the morphology of the samples. For this reason, SEM and EDS analyses were carried out next. From the backscattered electron images, the lamellar $SrCu_2(BO_3)_2$ matrix can be observed along with the presence of darker and lighter zones, which suggest the presence of impurities containing lighter and heavier elements, respectively. The lighter zone corresponds to the CuO impurities thus corroborating the PXRD results in **Figure 2**. The darker zones are attributed to the Mg-rich regions in the sample – more details on this can be found in the **Supplementary File, Figure S1**.

Representative EDS point analysis spectra collected on $SrCu_{2-x}Mg_x(BO_3)_2$ with $x$ = 0.03, 0.05 and 0.10 and the corresponding SEM images are summarised in **Figures 3a–c**. The line at 1.25 eV in the EDS spectrum is the hallmark of the Mg emission and should thus be taken as a semi-quantitative measure for the presence of Mg in the sample – **Table 1**. From the EDS point analysis, a clearly distinguished Mg emission peak is visible in all cases and thus unambiguously confirms the presence of Mg in the matrix for all three doping concentrations. Although the average semi-quantitative values yield an approximate $x$ = 0.07 for the nominal $x$ = 0.10 sample, a clear trend showing an increase in $x$ with the nominal doping concentrations is observed – see **Table 1**. The EDS point analysis also reveals a saturation of the Mg concentration that is successfully incorporated into the parent $SrCu_2(BO_3)_2$ structure, which is in qualitative agreement with the conclusions drawn from PXRD in **Figure 2b,c**.

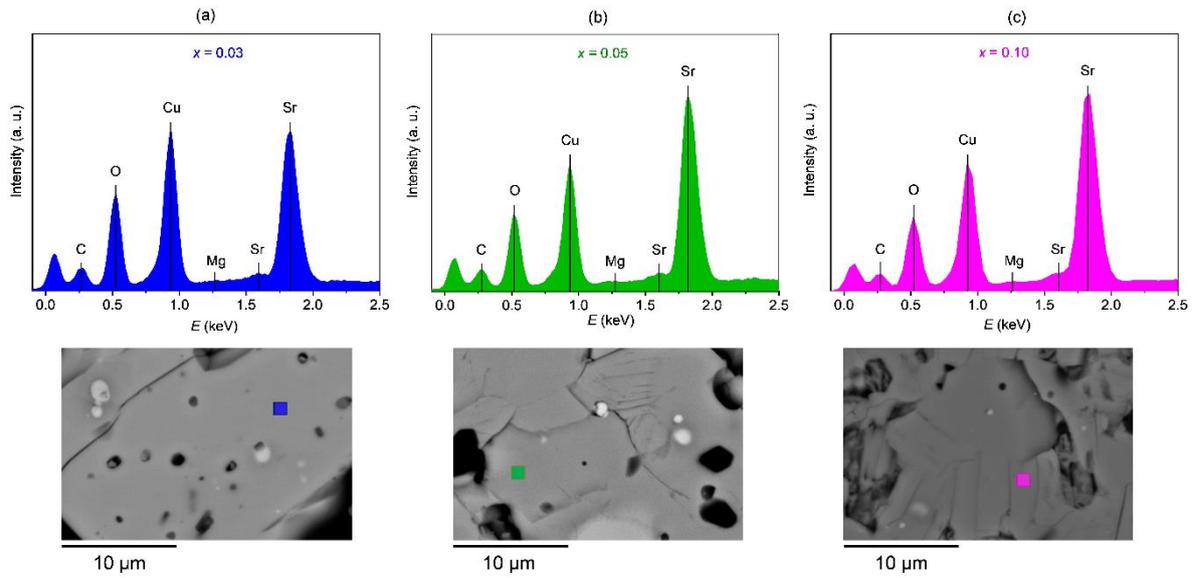

**Figure 3.** Representative EDS point analysis spectra collected on the matrix for $SrCu_{2-x}Mg_x(BO_3)_2$ with $x$ = 0.03 **(a)**, $x$ = 0.05 **(b),** and $x$ = 0.1 **(c)**. The corresponding SEM images can be seen below the EDS spectra. Coloured squares mark the regions which EDS spectra were collected from.

**Table 1**. The average semi-quantitative atomic % values of O, Mg, Cu and Sr together with the Mg/Cu, Mg/Sr and Cu/Sr ratios. These results were obtained from ~35 different SEM-EDS point analyses on $SrCu_{2-x}Mg_x(BO_3)_2$ samples with $x$ = 0.03, 0.05 and 0.10. The semi-quantifications were performed excluding boron as it was not possible to detect such a light element.

| | Sample, $SrCu_{2-x}Mg_x(BO_3)_2$ | | | | | |
|---|---|---|---|---|---|---|
| | $x$ = 0.03 | | $x$ = 0.05 | | $x$ = 0.10 | |
| Element | Experimental | Theoretical | Experimental | Theoretical | Experimental | Theoretical |
| | Concentration (at%) | | | | | |
| O | 63(2) | 66.67 | 57(7) | 66.67 | 57(7) | 66.67 |
| Mg | 0.4(1) | 0.33 | 0.6(1) | 0.54 | 0.8(3) | 1.11 |
| Cu | 24(2) | 21.89 | 27(4) | 21.67 | 28(4) | 21.11 |
| Sr | 12.6(4) | 11.11 | 15(3) | 11.11 | 14(3) | 11.11 |
| Sum, % | 100.0(2) | 100.0 | 99.6(4) | 100.0 | 99.8(3) | 100.0 |
| | | | | | | |
| Mg/Cu | 0.016 | 0.015 | 0.02 | 0.025 | 0.03 | 0.053 |
| Mg/Sr | 0.030 | 0.03 | 0.04 | 0.05 | 0.06 | 0.1 |
| Cu/Sr | 1.94 | 1.97 | 1.8 | 1.95 | 2.0 | 1.9 |
| | | | | | | |
| | Concentration, mol% | | | | | |
| Mg | 3.4(8) | 3.0 | 5(1) | 5.0 | 7(2) | 10.0 |

The presence of CuO and Mg-based secondary phase impurities observed in PXRD as well as in SEM stem from the partial incorporation of Mg into the parent $SrCu_2(BO_3)_2$ structure. The difficulty in reaching higher Mg concentrations in the $SrCu_2(BO_3)_2$ lattice could be linked to the low mobility of the $Mg^{2+}$ ions, which can be rationalised based on the fundamental diffusion mechanism and theory [58]. The atomic diffusivity depends on the composition, temperature and pressure. In our case, the low $SrCu_2(BO_3)_2$ decomposition temperature of ~ 950 °C limits the maximum annealing temperatures to 900 °C. At this temperature, $Mg^{2+}$ diffusion processes are slow, with diffusion constant values between $10^{-14}$–$10^{-11}$ m$^2$ s$^{-1}$ which severely limits the possibility for successful Mg-doping.

### 3.3. Magnetic susceptibility

Next, an investigation of the Mg-doping effects on the magnetic properties is presented. The magnetic susceptibilities, $\chi = M/H$, measured for undoped and Mg-doped $SrCu_2(BO_3)_2$ powders in the 2–300 K temperature range are compared in **Figure 4**. All samples generally show a qualitatively similar behaviour: at high temperatures, $\chi$ follows the Curie-Weiss law, reaching a maximum at around ~15 K, followed by a rapid suppression upon further cooling before it starts to increase again at the lowest temperatures. In order to quantitatively compare all samples, a fit of $\chi$ to a Curie-Weiss law in the high-temperature range of the data, between 100–300 K, is performed − **Equation 4**:

$$\chi = \frac{C}{T - \theta} + \chi_0 \quad . \tag{4}$$

Here $C$ denotes the Curie constant, $\theta$ is the Curie-Weiss temperature, while $\chi_0$ denotes a small temperature-independent diamagnetic contribution of the ion cores and/or the remnant signal of the sample holder. For the parent $SrCu_2(BO_3)_2$ sample, the fit to **Equation 4** yields a $\theta = -135(1)$ K, which is in excellent agreement with the literature data and can be described with the Shastry-Sutherland model given in **Equation 1**. For Mg-doped samples, a systematic and quite pronounced reduction in $\theta$ is observed – **Table 2**: $\theta$ decreases to $-129(1)$ K and then to $-110(1)$ K for 5% and 10% doped samples, respectively. Although one has to take these values with a grain of salt, the systematic reduction in $\theta$ with Mg-content implies a softening of the antiferromagnetic exchange interactions driven by the small structural changes observed in PXRD in **Figure 2**.

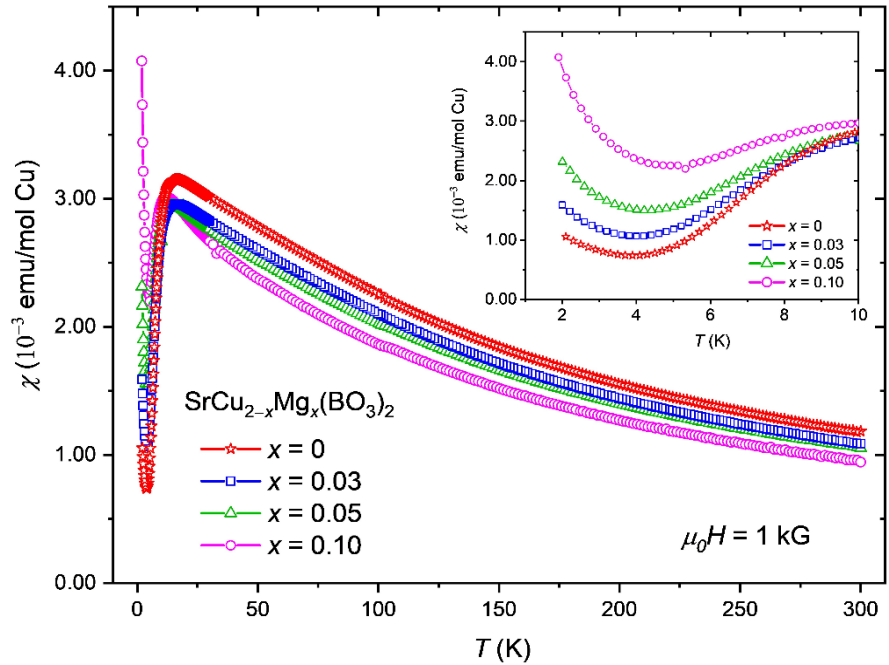

**Figure 4**. The magnetic susceptibility curves measured at $\mu_0 H = 1$ kG for $SrCu_{2-x}Mg_x(BO_3)_2$ with $x = 0, 0.03, 0.05$, and 0.1 powders. The inset shows the expanded low-temperature 2−10 K region where the upturn in the magnetic susceptibility can be clearly seen to increase with Mg concentration.

It can be further noticed that the temperature at which $\chi(T)$ has a maximum, $T_{max}$, also decreases from 16.6(2) K for the undoped sample to 12.2(1) K for the $x = 0.1$ sample. This parallels the reduction in $\theta$ discussed above and is also consistent with the softening of antiferromagnetic exchange interactions. Similarly, the introduction of nonmagnetic $Mg^{2+}$ ions into the lattice affects also the low-temperature magnetic susceptibility.

**Table 2**. The variation of spin gap, Δ, extracted from low-temperature susceptibility fits in the range 2–6 K, the $T_{max}$ which is the $\chi$ maximum temperature, the Curie constant C' for the low-temperature upturn in $\chi$ (for the optimal fit, a θ' = −0.5 K was used) and the estimated concentrations of unpaired $Cu^{2+}$ moments, and the Curie-Weiss temperature, $\theta$, extracted from Curie-Weiss fits of the susceptibility data in the temperature interval 100–300 K, for different Mg-doping concentrations. Comparison with undoped $SrCu_2(BO_3)_2$ is also made. See the main text for more details.

| Sample/$x$ | Δ [K] | $T_{max}$ [K] | C' (low-T, 1.8–3.5 K) in emu K/mol Cu | Estimated fraction of impurities/free Cu spins in % | Estimated Mg, in mol% | High-T fit, $\theta$ [K] |
|---|---|---|---|---|---|---|
| $SrCu_2(BO_3)_2$ $x = 0$ | 26.0(2) | 16.6(2) | 3.6(1) $10^{-3}$ | 0.95(1) | – | −135(1) |
| $SrCu_{1.97}Mg_{0.03}(BO_3)_2$ $x = 0.03$ | 22.6(2) | 16.2(3) | 5.2(2) $10^{-3}$ | 1.37(4) | 2.7(1) | −136(1) |
| $SrCu_{1.95}Mg_{0.05}(BO_3)_2$ $x = 0.05$ | 20.5(1) | 14.8(2) | 7.2(2) $10^{-3}$ | 1.9(1) | 3.9(1) | −129(1) |
| $SrCu_{1.9}Mg_{0.1}(BO_3)_2$ $x = 0.10$ | 17.1(3) | 12.2(1) | 10.7(2) $10^{-3}$ | 2.85(5) | 5.7(1) | −110(1) |

Magnetic susceptibility data, as shown in the inset of **Figure 4**, show a characteristic Curie upturn at low temperatures, which is a hallmark of unpaired $Cu^{2+}$ ions. In an attempt to quantitatively analyse $\chi(T)$ at low temperatures, we assume that it has three contributions [49] – a dominant thermally-activated contribution from the spin-dimer lattice characterised by the spin-gap Δ, a Curie-Weiss contribution from the unpaired $Cu^{2+}$ moments in the lattice introduced by Mg-doping and quantified by their Curie constant C', and the same temperature-independent contribution $\chi_0$ that was considered also in the high-temperature analysis:

$$\chi = \frac{C'}{T - \theta'} + a e^{\frac{-\Delta}{T}} + \chi_0 \quad . \tag{5}$$

In these fits, a very small Curie-Weiss temperature θ' = −0.5 K was fixed for optimal fitting. This minute θ' probably accounts for weak residual interactions between orphan $Cu^{2+}$ moments. Remarkably, just like $\theta$ and $T_{max}$, the effective spin gap also decreases with Mg-content from 26.0(2) K for $x = 0$ to just 17.1(3) K for the $x = 0.10$ sample, respectively. It should be stressed that the extracted Δ for the undoped sample matches the value from the literature where the same modelling was applied [49]. Moreover, for $x = 0.05$, a spin gap of 20.5(1) K is obtained which is in good agreement with the value reported in [47] for the same nominal doping concentration.

Although the presented analysis of $\chi$ for the spin-dimer phase may not yield exact values for the different magnetic parameters, the systematic decrease in $\theta$, $T_{max}$ as well as in Δ – **Table 2** – clearly points to a detrimental effect that Mg-doping has on the quantum magnetism of $SrCu_2(BO_3)_2$. The minute structural changes of the $Cu^{2+}$ dimer lattice in the Mg-doped samples – **Figure 2** – are probably not sufficient to adequately change the J'/J ratio and trigger the transition to a plaquette-singlet, quantum spin liquid or Néel states as the spin gap remains open for all Mg concentrations. Yet, these structural changes are sufficient to fine-tune the intradimer and/or interdimer exchange interactions and thus to affect extracted susceptibility parameters.

It is very intriguing that these changes occur concomitantly with the enhanced low-temperature Curie-upturn in $\chi$ (for the full analysis see the **Supplementary File, Table S2** and **S3**). This upturn cannot be attributed to any of the CuO and Mg-based secondary impurity phases detected in PXRD, as none of them is expected to show a similar paramagnetic contribution. We therefore conclude that the low-temperature upturn in $\chi$ is in fact intrinsic to Mg-doped $SrCu_2(BO_3)_2$ and is associated with the intrinsic lattice defects emerging when the nonmagnetic substitutional $Mg^{2+}$ ions break the Cu-spin dimers. This conclusion is further supported by low temperature X-band EPR data (*vide infra*).

In order to quantify the paramagnetic defects in the lattice introduced by Mg-doping, the low-temperature Curie constant $C'$ is calculated from Curie-Weiss fits in the temperature interval 1.8 K – 3.5 K and listed in **Table 2.** The individual fits are presented in the **Supplementary File, Figure S2**. A systematic increase in the $C'$ constant with Mg concentration is observed and thus the estimated fraction of impurities also increases with Mg concentration. Note that a similar Curie-upturn is found already in the undoped powder sample for which the calculated $Cu^{2+}$-impurity concentration corresponds to ~0.95%. Such concentration is comparable to the literature data for powder samples [36]. Moreover, for the nominal $x = 0.03$, the low-temperature upturn in doped samples is comparable with the literature [48]. However, for the highest $x = 0.05$ and 0.10, $C'$ significantly exceeds literature values thus implying that our approach to Mg doping allows for higher Mg-concentrations in the $SrCu_2(BO_3)_2$ lattice.

### 3.4. X-band Electron Paramagnetic Resonance Measurements (EPR)

In the next step, a local probe technique, X-band EPR is employed to investigate the intrinsic $Cu^{2+}$ defects that are created in the lattice upon Mg-doping. While EPR studies have been previously reported for undoped $SrCu_2(BO_3)_2$ in the form of single crystals or powders [20,21,22,23,24,25], there are, however, no reports on EPR spectroscopy studies on doped $SrCu_2(BO_3)_2$ yet. For Mg-doped samples, high temperature EPR spectra are still dominated by the well-known signal of the dimer lattice – **Supplementary File**, **Figure S3**. This signal broadens upon cooling due to the development of spin correlations and its intensity approximately mimics the temperature dependence of $\chi$. The main broad signal almost completely disappears below 10 K, i.e., when $T < \Delta$. At low temperatures, another signal with an axial $g$-factor anisotropy becomes visible – **Supplementary File**, **Figure S3**. While it is barely detectable for undoped $SrCu_2(BO_3)_2$, it becomes significantly more pronounced in Mg-doped samples. Its presence is especially evident in the $x = 0.10$ sample, where it dominates the spectrum already at 40 K as shown in **Figure 5a**, and can still be traced up to room temperature – **Supplementary File**, **Figure S3(d)**. The intensity of this EPR component increases with the decrease in temperature and with increasing Mg-doping concentration, as it is shown in **Supplementary File**, **Figure S4(d)**. This parallels the low-temperature dependence of the magnetic susceptibility and it is thus concluded that both contributions have the same origin – the intrinsic Cu moments from broken $Cu^{2+}$ dimers.

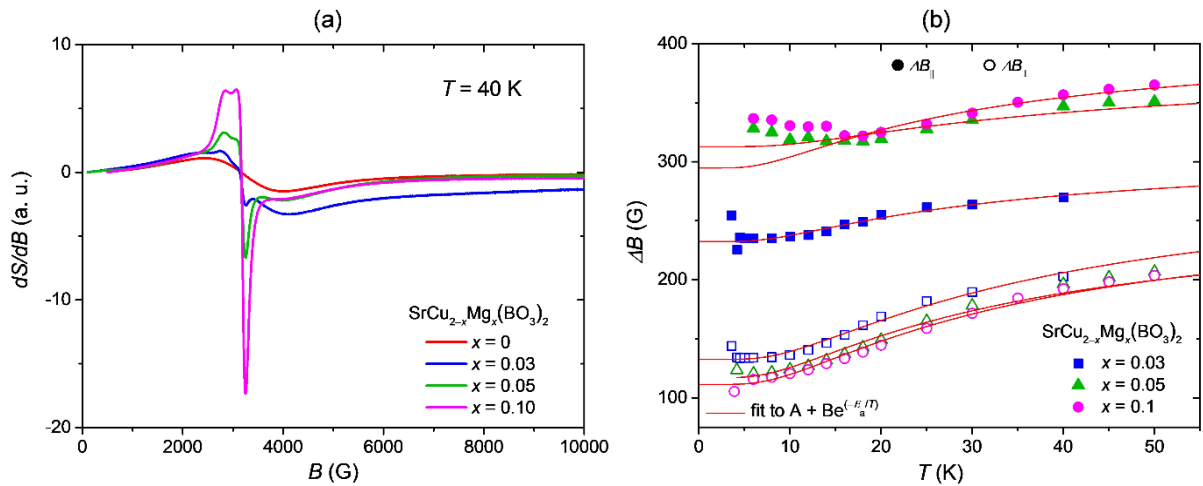

**Figure 5 (a)** The comparison of X-band EPR spectra of undoped $SrCu_2(BO_3)_2$ and Mg-doped $SrCu_2(BO_3)_2$ measured at 40 K. The EPR spectra of Mg-doped $SrCu_2(BO_3)_2$ powders clearly show the presence of intrinsic Cu moments from broken $Cu^{2+}$ dimers, represented by the sharp component with an axial g-factor anisotropy. **(b)** The temperature behaviour of X-band EPR linewidth data for low-temperature impurity peaks in the *ab*-plane, ($\Delta B_\perp$) and along the c-axis-plane ($\Delta B_\parallel$) for $SrCu_{2-x}Mg_x(BO_3)_2$ with $x = 0$, 0.03, 0.05, and 0.10. Solid lines are fit to the empirical $A + B \exp(-E_a/T)$ expression yielding a thermally activated energy $E_a \approx 30$ K for all sample compositions.

The EPR spectra are thus simulated to a sum of two components: a broad Lorentzian line with a large linewidth $\Delta B_{1/2}$ that considers the main signal of the dimer lattice and the low-temperature component with an axial g-factor anisotropy characterised by $g_{\parallel}$ and $g_{\perp}$ as the two g-factor eigen-values along the c-axis and in the ab-plane crystallographic directions. In the latter case, a similar uniaxial anisotropy in the linewidth with two values for the linewidth $\Delta B_{\parallel}$ and $\Delta B_{\perp}$, respectively, is also assumed. For the main component at $g$ = 2.095, $\Delta B_{1/2} \approx 1400$ G at room temperature is nearly identical for undoped as well as Mg-doped samples – **Supplementary File**, **Figure S4a**, and is consistent with the values determined for undoped $SrCu_2(BO_3)_2$ powders from the literature [20]. On cooling, $\Delta B_{1/2}$ initially monotonically increases in all samples, showing a similar temperature dependence that has also been previously reported for undoped $SrCu_2(BO_3)_2$. However, below 40 K, we find a more significant increase in $\Delta B_{1/2}$ for $x$ = 0.05 and 0.10 samples when compared to $x$ = 0 or undoped $SrCu_2(BO_3)_2$ – **Supplementary File**, **Figure S4a**. This indicates the emergence of internal fields in doped samples that provide additional broadening mechanism associated with the presence of intrinsic impurities. The low-temperature "impurity" EPR component can be described in all samples with similar EPR parameters: $g_{\parallel}$ = 2.38, $g_{\perp}$ = 2.008 – **Supplementary File, Figure S4c**, and a similarly pronounced linewidth anisotropy $\Delta B_{\parallel} \approx 320$ G and $\Delta B_{\perp} \approx 120$ G – **Figure 5b**. Below 50 K, $\Delta B_{\parallel}$ and $\Delta B_{\perp}$ start to decrease with the decreasing temperature. This temperature dependence can be empirically described by a thermally activated mechanism, $\Delta B \propto \exp(-E_a/T)$ – **Figure 5b**, with the activation energy $E_a \approx 30$ K for all cases. The extracted value for $E_a$ is very similar to the spin-gap values from spin susceptibility measurements, which suggests that these "impurity" spins are directly detecting the spin dynamics of the nearby unbroken Cu spin dimers. The spins contributing to this low-temperature signal are thus indeed intrinsic $Cu^{2+}$ defects from Mg-broken spin dimers sitting in the $[Cu(BO_3)]^-$ layers. Mg-doping is thus indeed successful and creates intrinsic $Cu^{2+}$ defects that couple to the parent dimer lattice and effectively probe its spin dynamics.

### 3.5. High Magnetic field measurements

Variations in the parameters extracted from $\chi$ and X-band EPR with the increasing concentration of intrinsic defects may influence the magnetisation plateaus and stabilise some novel states. Therefore, next, the magnetisation curves, $M(H)$, to fields up to 35 T for $SrCu_{2-x}Mg_x(BO_3)_2$ with $x$ = 0.03, 0.05 and 0.10 samples at $T$ = 0.4 K are reported. The onset of different phases that are stabilised as a function of magnetic field are best observed in d$M$/d$H$ plots – **Figure 6a**. In d$M$/d$H$, a plateau and a step appear as dip and peak, respectively. The dip can be seen as a kink when there is another increasing d$M$/d$H$ component. In **Figure 6a**, it can be seen that d$M$/d$H$ displays a peak at $\mu_0 H_{C4}$ = 29 T that corresponds to the onset of pseudo-1/8 plateau, analogous to the 1/8 plateau found in undoped $SrCu_2(BO_3)_2$ [36]. We also found a reduction of d$M$/d$H$ related to the 1/8 plateau phase in higher field site. With increasing Mg-doping, the peak slightly shifts towards lower fields and, most importantly, it is severely suppressed. This is consistent with expectations presented in [48] that the pseudo-1/$n$ plateaus would be suppressed with increasing doping. The anomalies in d$M$/d$H$ at $\mu_0 H'_{C3} \approx 25.0$ T and $\mu_0 H'_{C2} \approx 21.7$ T are less prominent in the present measurements, presumably because of the powder nature of the studied samples. According to [48], $H'_{C2}$ corresponds to localised bound states of triplets, whereas $H'_{C3}$ is the anomaly of the localised bound states with extra localised triplets in the vicinity. The bump in d$M$/d$H$ at $\mu_0 H'_{C1} \approx 17.1$ T is observed for all the Mg-doping concentrations. Following iPEPS calculations [48], this anomaly appears upon doping when two magnesium impurities break nearest-neighbouring dimers. The two liberated $Cu^{2+}$ $S$ = ½ spins couple to a singlet that converts to a triplet at $H'_{C1}$. Strong coupling of these pairs of $Cu^{2+}$ liberated spins renders them unobservable in X-band EPR spectra. Interestingly, for $x$ = 0.10, this anomaly appears the least pronounced. The most important result of these measurements is the observance of a clear maximum in d$M$/d$H$ at $\mu_0 H \approx 9$ T. The scaling of the d$M$/d$H$ maximum at $H'_{C0}$ with the intrinsic defect concentration determined from magnetic susceptibility and "impurity-peak" X-band EPR intensity suggests that liberated $Cu^{2+}$ moments must contribute to the stabilisation of novel impurity configurations. Previously, hints of an extra magnetic structure stabilised at this field were deduced from tunnel diode oscillator magnetic susceptibility measurements and were attributed to the stabilisation of the coupled $Cu^{2+}$ moments liberated by Mg-doping on the next-nearest (or even more distant) dimers [48]. Four examples of possible arrangements of pairs of liberated $Cu^{2+}$ $S$ = ½ spins are represented in **Figure 6b**. Due to the longer exchange pathways, the coupling of liberated $Cu^{2+}$ spins is weaker for these configurations and thus the gap to the excited triplet state closes at lower fields than in case of nearest-neighbouring liberated spins. The X-band EPR data discussed above may provide an extra insight into such next-nearest neighbouring configurations of coupled $Cu^{2+}$ spins. At elevated temperatures, liberated $Cu^{2+}$ spins relax mainly via nearby triplet excitations thermally activated across the spin gap $\Delta$ = 30 K – as a result, $\Delta B$ follows the thermally-activated temperature dependence as shown in **Figure 5b**. However, at temperatures well below $\Delta$, this broadening mechanism becomes less efficient and thus spin correlations between $Cu^{2+}$ moments on next-nearest broken dimers provide an additional broadening mechanism. This is especially pronounced in the samples with $x$ = 0.05 and 0.10, where $\Delta B_{\parallel}$ starts to increase again below ~15 K, which roughly corresponds to a critical field $H'_{c0}$. A combined high-field magnetisation study together with X-band EPR data therefore supports the picture of coupled $Cu^{2+}$ moments on nearest and next-nearest broken dimers as derived from iPEPS computations in [48].

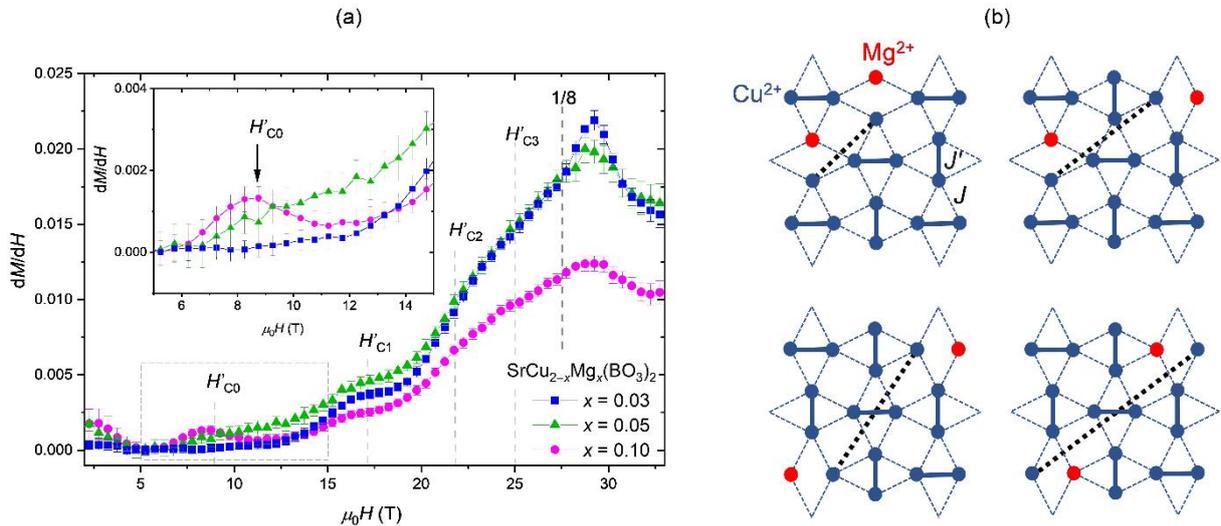

**Figure 6 (a)** High-field magnetic susceptibility measurements for $SrCu_{2-x}Mg_x(BO_3)_2$ with x = 0, 0.03. 0.05, and 0.10 powders in the region up to the pseudo-1/8 plateau. The critical field notations are termed after Ref. [48]. **(b)** Special 2-impurity configurations of dimer-free $Cu^{2+}$ $S$ = 1/2 spins remaining from Mg-magnetic dilution and coupled at distances beyond intra-dimer $Cu^{2+}$ bonds. The anomaly at $\mu_0 H'_{c0} \approx$ 9 T highlighted in the inset of **(a)** is due to such next-nearest $Cu^{2+}$ impurity pairs.

4. **CONCLUSIONS**

This work shows that it is possible to push the Mg-doping boundary to higher values and thus tackle the phase diagram of $SrCu_2(BO_3)_2$ over broader parameter space. It is demonstrated that by changing the source of Mg during the synthesis, the doping is pushed to an upper nominal limit of $x$ = 0.10 in $SrCu_{2-x}Mg_x(BO_3)_2$. This large concentration of Mg dopant has not been reported before, as it is notoriously difficult to introduce Mg impurities into $SrCu_2(BO_3)_2$ due to low $Mg^{2+}$ diffusion constants at applied reaction temperatures. At these higher Mg-doping levels more pronounced structural changes in Mg-doped samples are observed: the unit cell parameters and volume show a systematic increase with the increasing doping concentration. However, the incorporation of Mg in the $SrCu_2(BO_3)_2$ structure does not equal the nominal Mg concentration and even tends to saturate at lower values according to PXRD and SEM-EDS analyses. Nevertheless, the achieved Mg doping concentrations are already sufficient to affect the magnetic properties as expressed by a significant and systematic reduction of the spin gap, the Curie-Weiss temperature and the magnetic susceptibility maximum temperature with the increasing dopant concentration. These measurements are thus consistent with the breaking of $Cu^{2+}$ dimers by magnetic dilution and are further corroborated by X-band EPR spectroscopy. Finally, reaching higher $Mg^{2+}$ doping levels allows the detection of the $\mu_0 H'_{c0} \approx 9$ T anomaly more clearly. This anomaly is indicative of special configurations of dimer-free $Cu^{2+}$ $S = 1/2$ spins remaining from magnetic dilution and coupled at distances beyond the intra-dimer $Cu^{2+}$ bonds [48]. While the presented data on powder samples shows the potential of this synthesis approach, it also opens the next challenge of growing single crystals with incorporated $x = 0.10$ or higher concentration of Mg to perform a more in-debt study of the 9 T anomaly.

**ACKNOWLEDGEMENTS**

This project has received funding from the European Union's Horizon 2020 Research and Innovation Programme under the Marie Skłodowska-Curie grant agreement No. 101031415.
Support from the Slovenian Research and Innovation Agency (J2-2496, P2-0105, P2-0348, P1-0125), and the Jožef Stefan Institute Director's Fund, are gratefully acknowledged. Part of the research described in this paper was performed at the Canadian Light Source, a national research facility of the University of Saskatchewan, which is supported by the Canada Foundation for Innovation (CFI), the Natural Sciences and Engineering Research Council (NSERC), the National Research Council (NRC), the Canadian Institutes of Health Research (CIHR), the Government of Saskatchewan, and the University of Saskatchewan.
Maja Koblar is acknowledged for help with sample preparation for the SEM-EDS analyses. Assoc. Prof. Andrej Zorko, Asst. Prof. Matej Pregelj and Damjan Vengust are acknowledged for assistance with the setup for the EPR measurements. A part of this work was supported by GIMRT program of Institute for Materials Research, Tohoku University.

**REFERENCES**

[1] Balents, L. Spin liquids in frustrated magnets. *Nature* **2010**, *464*, 199–208. DOI: 10.1038/nature08917

[2] Broholm, C.; Cava, R. J.; Kivelson, S. A.; Nocera, D. G.; Norman, M. R.; Senthil, T. Quantum spin liquids. *Science* **2020**, *367*, 263. DOI: 10.1126/science.aay0668

[3] Anderson, P. W. The resonating valence bond state in $La_2CuO_4$ and superconductivity. *Science* **1987**, *235*, 1196–1198. DOI: 10.1126/science.235.4793.1196

[4] Vasiliev, A.; Volkova, O.; Zvereva, E.; Markina, M. Milestones of low-D quantum magnetism. *npj Quant. Mater*. **2018**, *3*, 18. DOI: 10.1038/s41535-018-0090-7

[5] Anderson, P. W. Resonating valence bonds: A new kind of insulator. *Mater. Res. Bull*. **1973**, *8*, 153–160. DOI: 10.1016/0025-5408(73)90167-0

[6] Mezzacapo, F. Ground-state phase diagram of the quantum $J_1 - J_2$ model on the square lattice. *Phys. Rev. B* **2012**, *86*, 45115. DOI: 10.1103/PhysRevB.85.060402

[7] Basov, D.; Averitt, R.; Hsieh, D. Towards properties on demand in quantum materials. *Nature Mater.* **2017**, *16*, 1077–1088. DOI: 10.1038/nmat5017

[8] Anderson, P. W. Sources of Quantum Protection in high-$T_C$ superconductivity. *Science* **2000**, *288*, 480–482. DOI: 10.1126/science.288.5465.480

[9] Taillefer, L. Superconductivity and quantum criticality. *Phys. Canada* **2011**, *67*, 109–112.

[10] Lee, P. A.; Nagaosa, N.; Wen, X.-G. Doping a Mott insulator: Physics of high-temperature superconductivity. *Rev. Mod. Phys*. **2006**, *78*, 17. DOI: 10.1103/RevModPhys.78.17

[11] Azuma, M.; Hiroi, Z.; Takano, M.; Ishida, K.; Kitaoka, Y. Observation of a Spin Gap in $SrCu_2O_3$ Comprising Spin-½ Quasi-1D Two-Leg Ladders. *Phys. Rev. Lett*. **1994**, *73*, 2626. DOI: 10.1103/PhysRevLett.73.3463

[12] Barnes, T.; Riera, J. Susceptibility and excitation spectrum of $(VO)_2P_2O_7$ in ladder and dimer-chain models. *Phys. Rev. B* **1994**, *50,* 6817–6822. DOI: 10.1103/physrevb.50.6817. PMID: 9974635.

[13] Tsirlin, A. A; Shpanchenko, R. V.; Antipov, E. V.; Bougerol, C.; Hadermann, J.; Van Tendeloo, G.; Schnelle, W.; Rosner, H. Spin ladder compound $Pb_{0.55}Cd_{0.45}V_2O_5$: Synthesis and investigation. *Phys. Rev. B* **2007**, *76*, 104429. DOI: 10.1103/PhysRevB.76.104429

[14] Kodama, K.; Fukamachi, T.; Harashina, H.; Kanada, M.; Kobayashi, Y.; Kasai, M.; Sasaki, H.; Sato, M.; Kakurai K. Spin-Gap Behavior of $CuNb_2O_6$. *J. Phys. Soc. Jpn*. **1998**, *67*, 57–60. DOI: 10.1143/JPSJ.67.57

[15] Dagotto, E.; Rice, T. M. Surprises on the Way from One-to Two-Dimensional Quantum Magnets: The Ladder Materials. *Science* **1996**, *271*, 618–623. DOI: 10.1126/science.271.5249.618

[16] Taniguchi, S.; Noshikawa, T.; Y. Yasui, Y.; Kobayashi, Y.; Sato, M.; Nishioka, T.; Kontani, M.; Sano, K. Spin gap behavior of *S* = 1/2 quasi-two-dimensional system $CaV_4O_9$. *J. Phys. Soc. Jpn*. **1995**, *64*, 2758–2761. DOI: 10.1143/JPSJ.64.2758

[17] Kodama, K.; Harashina, H.; Shamot, S. Evolution of Spin Gap in the Excitation Spectra of Quasi-Two-Dimensional *S* = 1/2 System $CaV_4O_9$. *J. Phys. Soc. Jpn*. **1996**, *65*, 1941–1944. DOI: 10.1143/JPSJ.65.1941

[18] Kageyama, H. New Two-Dimensional Spin Gap Material $SrCu_2(BO_3)_2$. *J. Phys. Soc. Jpn*. **2000**, *69*, 65–71.

[19] Shastry, B. S.; Sutherland, B. Exact ground state of a quantum mechanical antiferromagnet. *Phys. B+C* **1981**, *108*, 1069–1070. DOI: 10.1016/0378-4363(81)90838-X

[20] Zorko, A.; Arčon, D.; Lappas, A.; Giapintzakis, J. Near critical behavior in the two-dimensional spin-gap system $SrCu_2(BO_3)_2$. *Phys. Rev. B* **2001**, *65*, 024417−024422. DOI: 10.1103/PhysRevB.65.024417

[21] Nojiri, H.; Kageyama, H.; Ueda, Y.; Motokawa, M. ESR Study on the Excited State Energy Spectrum of $SrCu_2(BO_3)_2$—A Central Role of Multiple-Triplet Bound States. *J. Phys. Soc. Jpn*. **2003**, *72*, 3243–3253. DOI: 10.1143/JPSJ.72.3243

[22] Zorko, A. Study of one- and two-dimensional magnetic systems with spin-singlet ground state. Ph.D. dissertation, University of Ljubljana, Slovenia **2004**. Accessed on 18[th] September 2023.

[23] Zorko, A.; Arčon, D.; Nuttall, C. J.; Lappas, A. X-band ESR study of the 2D spin-gap system $SrCu_2(BO_3)_2$. *J. Magn. Magn. Mater.* **2004**, *272*, 699–701. DOI: 10.1016/j.jmmm.2003.12.336

[24] Cépas, O; Kakurai, K.; Regnault, L. P.; Ziman, T.; Boucher, J. P.; Aso, N.; Nishi, M.; Kageyama, H.; Ueda, Y. Dzyaloshinski-Moriya Interaction in the 2D Spin Gap System $SrCu_2(BO_3)_2$. *Phys. Rev. Lett*. **2001**, *87*, 167205. DOI: 10.1103/PhysRevLett.87.167205

[25] Zorko, A.; Arčon, D.; Kageyama, H.; Lappas, A. Magnetic anisotropy of the $SrCu_2(BO_3)_2$ system as revealed by X-band ESR. *Appl. Magn. Reson.* **2004**, *27*, 267–278. DOI: 10.1007/BF03166320

[26] Takigawa, M.; Matsubara, S.; Horvatić, M.; Berthier, C.; Kageyama, H.; Ueda, Y. NMR Evidence for the Persistence of a Spin Superlattice Beyond the 1/8 Magnetization Plateau in $SrCu_2(BO_3)_2$. *Phys. Rev. Lett.* **2008**, *101*, 037202. DOI: 1103/physrevlett.101.037202

[27] Waki, T.; Arai, K.; Takigawa, M.; Saiga, Y.; Uwatoko, Y.; Kageyama, H.; Ueda, Y. A Novel Ordered Phase in $SrCu_2(BO_3)_2$ under High Pressure. *J. Phys. Soc. Jpn.* **2007**, *76*, 073710–073714. DOI: 10.1143/JPSJ.76.073710

[28] Cui, Y.; Liu, L.; Lin, H.; Wu, K.-H.; Hong, W.; Liu, X.; Li, C.; Hu, Z.; Xi, N.; Li, S.; Yu, R.; Sandvik, A. W.; Yu, W. Proximate deconfined quantum, critical point in $SrCu_2(BO_3)_2$. *Science* **2023**, *380*, 1179–1184. DOI: 10.1126/science.adc9487

[29] Kohlrautz, J.; Haase, J.; Green, E. L.; Zhang, Z. T.; Wosnitza, J.; Herrmannsdörfer, T.; Dabkowska, H. A.; Gaulin, B. D.; Stern, R.; Kühne, H. Field-stepped broadband NMR in pulsed magnets and application to SrCu$_2$(BO$_3$)$_2$ at 54 T. *J. Magn. Reson.* **2016**, *271*, 52–59. DOI: 10.1016/j.jmr.2016.08.005

[30] Koga, A.; Kawakami, N. Quantum Phase Transitions in the Shastry-Sutherland Model for SrCu$_2$(BO$_3$)$_2$. *Phys. Rev. Lett.* **2000**, *84*, 4461–4461. DOI: 10.1103/PhysRevLett.84.4461

[31] Kodama, K.; Yamazaki, J.; Takigawa, M.; Kageyama, H.; Onizuka, K.; Ueda, Y. Cu nuclear spin-spin coupling in the dimer singler state in SrCu$_2$(BO$_3$)$_2$. *J. Phys.: Condens. Matter* **2002**, *14*, L319. DOI: 10.1088/0953-8984/14/17/101

[32] Yang, J.; Sandvik, A. W.; Wang, L. Quantum criticality and spin liquid phase in the Shastry Sutherland model. *Phys. Rev. B*. **2022**, *105*, L060409. DOI: 10.1103/PhysRevB.105.L060409

[33] Miyahara, S.; Ueda, K. Exact Dimer Ground State of the Two Dimensional Heisenberg Spin System SrCu$_2$(BO$_3$)$_2$. *Phys. Rev. Lett.* **1999**, *82*, 3701–3704. DOI: 10.1103/PhysRevLett.82.3701

[34] Jaime, M.; Daou, R.; Crooker, S. A.; Weickert, F.; Uchida, A.; Feiguin, A. E.; Batista, C. D.; Dabkowska, H. A.; Gaulin, B. D. Magnetostriction and magnetic texture to 100.75 Tesla in frustrated SrCu$_2$(BO$_3$)$_2$. *Proc. Natl. Acad. Sci. U. S. A*. **2012**, *109*, 12404–12407. DOI: 10.1073/pnas.1200743109

[35] Matsuda, Y. H.; Abe, N.; Takeyama, S.; Kageyama, H.; Corboz, P.; Honecker, A.; Manmana, S. R.; Foltin, G. R.; Schmidt, K.P.; Mila, F. Magnetization of SrCu$_2$(BO$_3$)$_2$ in ultrahigh magnetic fields up to 118 T. *Phys. Rev. Lett.* **2013**, *111*, 137204. DOI: 10.1103/PhysRevLett.111.137204

[36] Kageyama, H.; Yoshimura, K.; Stern, R.; Stern, Mushnikov, N. V.; Onizuka, K.; Kato, M.; Kosuge, K.; Slichter, C. P.; Goto, T.; Ueda Y. Exact Dimer Ground State and Quantized Magnetization Plateaus in the Two-Dimensional Spin System SrCu$_2$(BO$_3$)$_2$. *Phys. Rev. Lett.* **1999**, *82*, 3168–3171. DOI: 10.1103/PhysRevLett.82.3168

[37] Nojiri, H.; Kageyama, H.; Onizuka, K.; Ueda, Y.; Motokawa, M. Direct Observation of the Multiple Spin Gap Excitations in Two-Dimensional Dimer System SrCu$_2$(BO$_3$)$_2$. *J. Phys. Soc. Jpn.* **1999**, *68*, 2906–2909. DOI: 10.1143/JPSJ.68.2906

[38] Kodama, K.; Takigawa, M.; Horvatic, M.; Berthier, C.; Kageyama, H.; Ueda, Y.; Miyahara, S.; Becca, F.; Mila, F. Magnetic Superstructure in the Two-dimensional Quantum Antiferromagnet SrCu$_2$(BO$_3$)$_2$. *Science* **2002**, *298,* 395–398. DOI: 10.1126/science.1075045

[39] Onizuka, K.; Kageyama, H.; Narumi, Y.; Kindo, K.; Ueda, Y.; Goto, T. 1/3 Magnetization Plateau in SrCu$_2$(BO$_3$)$_2$ - Stripe Order of Excited Triplets -. *J. Phys. Soc. Japan* 2000, *69*, 1016.

[40] Sebastian, S.; Harrison, N.; Sengupta, P.; Batista, C. D.; Francoual, S.; Palm, E.; Murphy, T.; Marcano, N.; Dabkowska, H. A.; Gaulin, B. D. Fractalization drives crystalline states in a frustrated spin system. *Proc. Natl. Acad. Sci. U. S. A.* **2008**, *105*, 20157–20160. DOI: 10.1073/pnas.0804320105

[41] Fogh, E.; Nayak, M.; Prokhnenko, O.; Bartkowiak, M.; Munakata, K.; Soh, J.-R.; Turrini, A. A.; Zayed, M. E.; Pomjakushina, E.; Kageyama, H.; Nojiri, H.; Kakurai, K.; Normand, B.; Mila, F.; Rønnow, H. M. Field-induced bound-state condensation and spin-nematic phase in SrCu$_2$(BO$_3$)$_2$ revealed by neutron scattering up to 25.9 T. *Nat. Commun.* **2024**, *15*, 442. DOI: 10.1038/s41467-023-44115-z

[42] Haravifard, S.; Graf, D.; Feiguin, A. E.; Batista, C. D.; Lang, J. C.; Silevitch, D. M.; Srajer, G.; Gaulin, B. D.; Dabkowska, H. A.; Rosenbaum, T. F. Crystallization of spin superlattices with pressure and field in the layered magnet SrCu$_2$(BO$_3$)$_2$. *Nat. Commun.* **2016**, *20*, 11956–11961. DOI: 10.1038/ncomms11956

[43] Sakurai, T.; Hirao, Y.; Hijii, K.; Okubo, S.; Ohta, H.; Uwatoko, Y.; Kudo, K.; Koike, Y. Direct Observation of the Quantum Phase Transition of SrCu$_2$(BO$_3$)$_2$ by High-Pressure and Terahertz Electron Spin Resonance. *J. Phys. Soc. Jpn.* **2008**, *87*, 033701–033705. DOI: 10.7566/JPSJ.87.033701

[44] Guo, J.; Sun, G.; Zhao, B.; Wang, L.; Hong, W.; Sidorov, V. A.; Ma, N.; Wu, Q.; Li, S.; Meng, Z. Y.; Sandvik, A. W.; Sun, L. Quantum phases of SrCu$_2$(BO$_3$)$_2$ from high-pressure thermodynamics. *Phys. Rev. Lett.* **2020**, *124*, 206602–206608. DOI: 10.1103/PhysRevLett.124.206602

[45] Shastry, B. S.; Kumar, B. SrCu$_2$(BO$_3$)$_2$: A unique Mott Hubbard insulator. *Prog. Theor. Phys. Suppl.* **2002**, *145*, 1–16. DOI: 10.1143/PTPS.145.1

[46] Yang, B.-J.; Kim, Y. B.; Yu, J.; Park, K. Doped valence-bond solid and superconductivity on the Shastry-Sutherland lattice. *Phys. Rev. B* **2008**, *77*, 104507. DOI: 10.1103/PhysRevB.77.104507

[47] Aczel, A. A.; MacDougall, G. J.; Rodriguez, J. A., Luke, G. M.; Russo, P. L.; Savici, A. T.; Uemura, Y. J.; Dabkowska, H. A.; Wiebe, C. R.; Janik, J. A.; Kageyama, H. Impurity-induced singlet breaking in SrCu$_2$(BO$_3$)$_2$. *Phys. Rev. B* **2007**, *76*, 214427–214433. DOI: 10.1103/PhysRevB.76.214427

[48] Shi, Z.; Steinhardt, W.; Graf, D.; Corboz, P.; Weickert, F.; Harrison, N.; Jaime, M.; Marjerrison, C.; Dabkowska, H. A.; Mila, F.; Haravifard, S. Emergent bound states and impurity pairs in chemically doped Shastry-Sutherland system. *Nat. Comm.* **2019**, *10*, 1–9. DOI: 10.1038/s41467-019-10410-x

[49] Liu, G. T.; Luo, J. L.; Wang, N. L.; Jing, X. N.; Jin, D.; Xiang, T.; Wu, Z. H. Doping effects on the two-dimensional spin dimer compound SrCu$_2$(BO$_3$)$_2$. *Phys. Rev. B* **2005**, *71*, 014441. DOI: 10.1103/PhysRevB.71.014441

[50] Dabkowska, H. A.; Dabkowski, A. B.; Luke, G. M., Dunsiger, S. R.; Haravifard, S.; Cecchinel, M.; Gaulin, B. D. Crystal growth and magnetic behaviour of pure and doped SrCu$_2$($^{11}$BO$_3$)$_2$. *J. Cryst. Growth* **2007**, *306*, 123–128. DOI: 10.1016/j.jcrysgro.2007.04.040

[51] Haravifard, S.; Dunsinger, S. R.; El Shawish, S.; Gaulin, B. D.; Dabkowska, H. A.; Telling, M. T. F.; Perring, T. G.; Bonča, J. In-gap Spin Excitations and Finite Triplet Lifetimes in the Dilute Singlet Ground State System SrCu$_{2-x}$Mg$_x$(BO$_3$)$_2$. *Phys. Rev. Let.* **2006**, *97*, 247206. DOI: 10.1103/PhysRevLett.97.247206

[52] Liu, G. T.; Luo, J. L.; Guo, Y. Q.; Su, S. K.; Zheng, P.; Wang, N. L.; Jin, D.; Xiang, T. In-plane substitution effect on the magnetic properties of the two-dimensional spin-gap system SrCu$_2$(BO$_3$)$_2$. *Phys. Rev. B*. **2006**, *73*, 014414. DOI: 10.1103/PhysRevB.73.014414

[53] Shannon, R. D. Revised Effective Ionic Radii and Systematic Studies of Interatomic Distances in Halides and Chalcogenides. *Acta Cryst.* **1976**, *A32,* 751–767. DOI: 10.1107/S0567739476001551

[54] Dragomir, M.; Šibav, L. Synthesis peculiarities of SrCu$_2$(BO$_3$)$_2$. *Manuscript in preparation* **2024**.

[55] Toby, B. H.; Von Dreele, R. B. GSAS-II: the genesis of a modern open-source all purpose crystallography software package. *J. Appl. Crystallogr.* **2013**, *46*, 544–549. DOI: 10.1107/S0021889813003531

[56] Smith, R. W.; Keszler, D. A. Synthesis, structure, and properties of the orthoborate SrCu$_2$(BO$_3$)$_2$. *J. Solid State Chem.* **1991**, *93*, 430–435. DOI: 10.1016/0022-4596(91)90316-A

[57] Goodenough, J. B. Magnetism and the Chemical Bond, Interscience Publisher, New York **1963**.

[58] Van Orman, J. A.; Crispin, K. L. Diffusion in Oxides. *Rev. Mineral. Geochem.* **2010**, *72*, 757−825. DOI: 10.2138/rmg.2010.72.17

# Supplementary Information File

**Higher-magnesium-doping effects on the singlet ground state of the Shastry-Sutherland SrCu$_2$(BO$_3$)$_2$**

Lia Šibav[1,2], Žiga Gosar[1,3], Tilen Knaflič[1,4], Zvonko Jagličić[5,6], Graham King[7], Hiroyuki Nojiri[8], Denis Arčon[1,3] and Mirela Dragomir[1,2*]

[1]Jožef Stefan Institute, Jamova cesta 39, 1000 Ljubljana, Slovenia
[2]Jožef Stefan International Postgraduate School, Jamova cesta 39, 1000 Ljubljana, Slovenia
[3]Faculty of Mathematics and Physics, University of Ljubljana, Jadranska ulica 19, 1000 Ljubljana, Slovenia
[4]Institute for the Protection of Cultural Heritage of Slovenia, Research Institute, Poljanska cesta 40, 1000 Ljubljana, Slovenia
[5]Institute of Mathematics, Physics and Mechanics, Jadranska ulica 19, 1000 Ljubljana, Slovenia
[6]Faculty of Civil and Geodetic Engineering, University of Ljubljana, Jamova cesta 2, 1000 Ljubljana, Slovenia
[7]Canadian Light Source, 44 Innovation Blvd, Saskatoon, SK S7N 2V3, Canada
[8] Institute for Materials Research, Tohoku University, Katahira 2-1-1, Sendai, 980-8577 Japan

*Corresponding author: mirela.dragomir@ijs.si

**Table S1.** The unit cell parameters and the unit cell volume for different Mg-doping concentrations as obtained from Rietveld refinement analysis. A systematic increase in the unit cell parameter $a$, $b$ and $c$ values as well as in the unit cell volume is observed with increasing doping concentrations.

|  | $a$, $b$ (Å) | $c$ (Å) | $V$ (Å$^3$) |
|---|---|---|---|
| SrCu$_2$(BO$_3$)$_2$ $x = 0.0$ | 8.9900(2) | 6.6483(2) | 537.32(3) |
| SrCu$_{1.97}$Mg$_{0.03}$(BO$_3$)$_2$ $x = 0.03$ | 8.9938(2) | 6.6534(2) | 538.18(4) |
| SrCu$_{1.95}$Mg$_{0.05}$(BO$_3$)$_2$ $x = 0.05$ | 8.9957(2) | 6.6562(2) | 538.64(2) |
| SrCu$_{1.9}$Mg$_{0.1}$(BO$_3$)$_2$ $x = 0.10$ | 8.9978(5) | 6.6584(5) | 539.08(8) |

**Table S2.** The complete results of magnetic susceptibility data fits in the temperature interval 2–6 K for different doping concentrations using the equation for total susceptibility at low temperatures (Equation 5 in the main text). Due to a high correlation between fitted parameters $C'$ and $\theta'$, the fitting was performed using multiple fixed $\theta'$ values between 0 to −1, assuming antiferromagnetic interactions between the intrinsic impurities − dimer-free Cu$^{2+}$ ions.

|  | $C'$ (emu K/g) | $\chi_0$ (emu/g) | $\theta'$ | $\Delta$ (K) | $a$ (emu/g) |
|---|---|---|---|---|---|
| SrCu$_2$(BO$_3$)$_2$ $x = 0$ | 1.04(1) · 10$^{-5}$ | 1.04(4) · 10$^{-6}$ | 0 | 27.9(3) | 4.7(2) · 10$^{-4}$ |
|  | 1.473(1) · 10$^{-5}$ | 2.73(3) · 10$^{-7}$ | −0.5 | 26.61(2) | 4.03(1) · 10$^{-4}$ |
|  | 2.01(1) · 10$^{-5}$ | −5.5(3) · 10$^{-7}$ | −1 | 25.6(2) | 3.57(8) · 10$^{-4}$ |
| SrCu$_{1.97}$Mg$_{0.03}$(BO$_3$)$_2$ $x = 0.03$ | 1.520(9) · 10$^{-5}$ | 1.65(4) · 10$^{-6}$ | 0 | 24.1(2) | 2.57(9) · 10$^{-4}$ |
|  | 2.18(1) · 10$^{-5}$ | 4.5(4) · 10$^{-7}$ | −0.5 | 22.6(2) | 2.15(5) · 10$^{-4}$ |
|  | 2.98(2) · 10$^{-5}$ | −7.8(6) · 10$^{-7}$ | −1 | 21.5(2) | 1.91(6) · 10$^{-4}$ |
| SrCu$_{1.95}$Mg$_{0.05}$(BO$_3$)$_2$ $x = 0.05$ | 2.24(1) · 10$^{-5}$ | 2.62(5) · 10$^{-6}$ | 0 | 22.8(3) | 1.98(9) · 10$^{-4}$ |
|  | 3.26(1) · 10$^{-5}$ | 7.1(4) · 10$^{-7}$ | −0.5 | 20.5(1) | 1.53(3) · 10$^{-4}$ |
|  | 4.51(2) · 10$^{-5}$ | −1.31(7) · 10$^{-6}$ | −1 | 18.9(2) | 1.32(3) · 10$^{-4}$ |
| SrCu$_{1.9}$Mg$_{0.1}$(BO$_3$)$_2$ $x = 0.10$ | 3.962(9) · 10$^{-5}$ | 3.33(4) · 10$^{-6}$ | 0 | 21.0(2) | 1.36(4) · 10$^{-4}$ |
|  | 4.84(2) · 10$^{-5}$ | 1.56(7) · 10$^{-6}$ | −0.25 | 18.8(2) | 1.08(4) · 10$^{-4}$ |
|  | 5.84(4) · 10$^{-5}$ | −3(1) · 10$^{-7}$ | −0.5 | 17.1(3) | 9.3(4) · 10$^{-5}$ |
|  | 6.96(6) · 10$^{-5}$ | −2.3(2) · 10$^{-6}$ | −0.75 | 15.9(4) | 8.5(4) · 10$^{-5}$ |
|  | 8.20(7) · 10$^{-5}$ | −4.4(2) · 10$^{-6}$ | −1 | 15.1(3) | 8.2(3) · 10$^{-5}$ |

**Table S3.** The complete results of high-temperature magnetic susceptibility data fits in the interval 100–300 K for different doping concentrations using the Curie-Weiss law equation (Equation 4 in the main text).

|  | $C$ (emu K/g) | $\theta$ (K) | $\chi_0$ (emu/g) |
|---|---|---|---|
| SrCu$_2$(BO$_3$)$_2$ | 3.32(2) · 10$^{-5}$ | −135(1) | 1.2806(4) · 10$^{-4}$ |
| SrCu$_{1.97}$Mg$_{0.03}$(BO$_3$)$_2$ | 3.14(2) · 10$^{-5}$ | −134.2(7) | 1.2793(2) · 10$^{-4}$ |
| SrCu$_{1.95}$Mg$_{0.05}$(BO$_3$)$_2$ | 2.90(2) · 10$^{-5}$ | −128.8(8) | 1.2823(2) · 10$^{-4}$ |
| SrCu$_{1.9}$Mg$_{0.1}$(BO$_3$)$_2$ | 2.33(2) · 10$^{-5}$ | −110.1(9) | 1.2873(2) · 10$^{-4}$ |

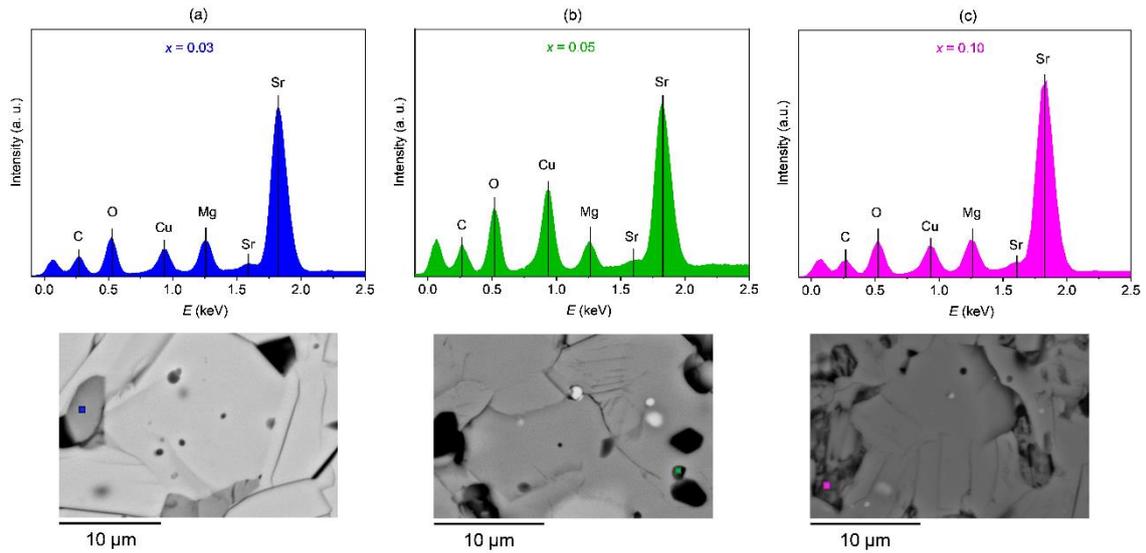

**Figure S1.** EDS point analysis spectra collected on the impurity sites for SrCu$_{2-x}$Mg$_x$(BO$_3$)$_2$, with $x$ = 0.03 **(a)**, $x$ = 0.05 **(b)** and $x$ = 0.10 **(c)**. A higher concentration of Mg is suggested from the impurity sites compared to the doped SrCu$_2$(BO$_3$)$_2$ matrix. The corresponding SEM images can be seen below the EDS spectra.

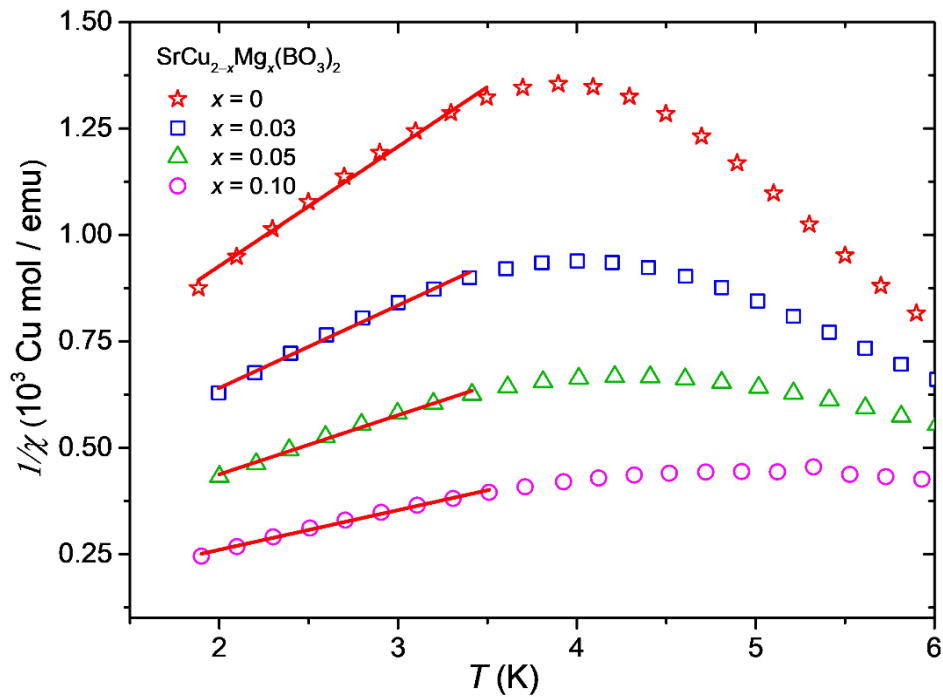

**Figure S2.** The low-temperature fits of the 1/$\chi$ data to Equation 4 (main text) at $T$ = 1.8–3.5 K for undoped and doped SrCu$_{2-x}$Mg$_x$(BO$_3$)$_2$ samples with $x$ = 0, 0.03, 0.05, and 0.10. The concentration of Cu$^{2+}$ $S$ = 1/2 impurities was extracted from these fits for all doping concentrations.

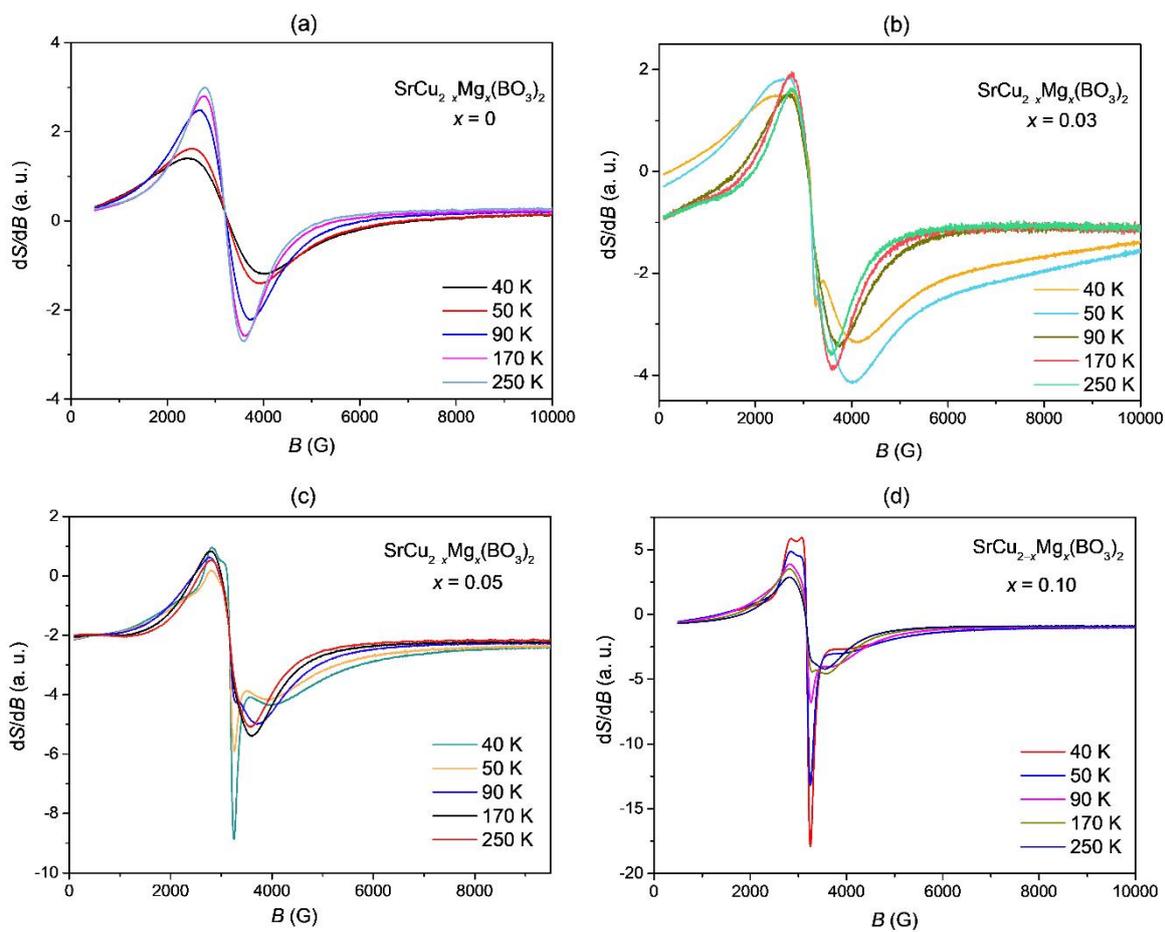

**Figure S3.** The EPR spectra of undoped SrCu$_2$(BO$_3$)$_2$ or $x$ = 0 **(a)** and SrCu$_{2-x}$Mg$_x$(BO$_3$)$_2$ with $x$ = 0.03 **(b)**, $x$ = 0.05 **(c)**, and $x$ = 0.10 **(d)** measured at temperatures of 40, 50, 90, 170 and 250 K, which show the development of low-temperature impurity peak that increases in intensity with increasing Mg-doping concentration.

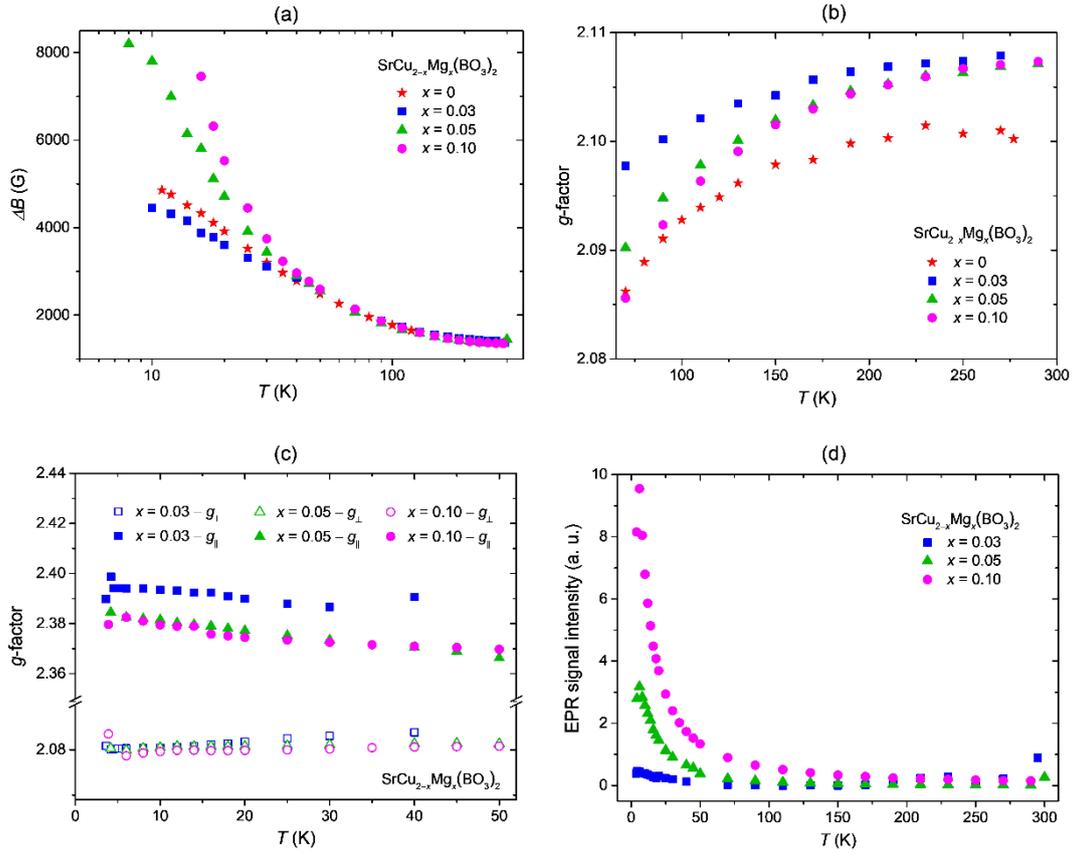

**Figure S4. (a)** The logarithmic temperature behaviour of the X-band EPR peak-to-peak linewidth of the main dimer-lattice signal. **(b)** The temperature behaviour of the $g$-factor of the main dimer-lattice signal in SrCu$_{2-x}$Mg$_x$(BO$_3$)$_2$ with $x$ = 0, 0.03, 0.05, and 0.10. **(c)** The temperature behaviour of the X-band EPR $g$-factors $g_\parallel$ and $g_\perp$ as the two $g$-factor values along the $c$ and $ab$-plane crystallographic directions for low-temperature "impurity" peaks for SrCu$_{2-x}$Mg$_x$(BO$_3$)$_2$ with $x$ = 0, 0.03, 0.05 and 0.10. **(d)** The temperature dependence of the X-band EPR signal intensity for low-temperature "impurity" peaks for $x$ = 0, 0.03, 0.05 and 0.10 samples.

5